\documentclass[a4paper,3p,sort&compress]{elsarticle}

\usepackage{booktabs}
\usepackage{graphicx}
\usepackage{xspace}
\usepackage{booktabs}
\usepackage[draft,inline,nomargin]{fixme}
\usepackage{makecell} 
\usepackage{lineno}
\usepackage{natbib}
\usepackage{amsmath}

\journal{ArXiv}

\begin{document}
%\linenumbers

% Macro para escribir NO$_2$
\newcommand{\no}{NO\textsubscript{2}\xspace}

\begin{frontmatter}

  \title{Probabilistic forecasting approaches for extreme \no episodes: a comparison of models}

% \author{} 
% \address{Artificial Intelligence Department\\Universidad Nacional de
%   Educaci\'on a Distancia --- UNED\\c/ Juan del Rosal, 16, Madrid, Spain}

  \author{Sebasti\'an P\'erez Vasseur}
  \author{Jos\'e L. Aznarte}
  \address{Artificial Intelligence Department\\Universidad Nacional de
    Educaci\'on a Distancia --- UNED\\c/ Juan del Rosal, 16, Madrid, Spain}
  \ead{jlaznarte@dia.uned.es}

\begin{abstract}
  % Air quality is an increasingly alarming issue in cities around the
  % world and thus there is an urgent need to reliably forecast high
  % concentration episodes of certain pollutants in the air. In this
  % sense, \no is one of the most worrisome pollutants, as it has been
  % proven its direct implication in a variety of serious health
  % affections. In some cities, high concentration episodes for \no are
  % dealt with by authorities through traffic restriction measures,
  % amongst others, which are activated when air quality deteriorates
  % beyond certain thresholds.

High concentration episodes for NO2 are increasingly dealt with by authorities
through traffic restrictions which are activated when air quality deteriorates
beyond certain thresholds. Foreseeing the probability that pollutant
concentrations reach those thresholds becomes thus a necessity.

Probabilistic forecasting is a family of techniques that allow for the
prediction of the expected distribution function instead of a single value. In
the case of NO2, it allows for the calculation of future chances of exceeding
thresholds and to detect pollution peaks.

We thoroughly compared 10 state of the art probabilistic predictive models,
using them to predict the distribution of NO2 concentrations in a urban location
for a set of forecasting horizons (up to 60 hours). Quantile gradient boosted
trees shows the best performance, yielding the best results for both the
expected value and the forecast full distribution. Furthermore, we show how this
approach can be used to detect pollution peaks.

% Episodic high concentrations of \no is an alarming issue in cities
%   around the world, and is usually dealt with by authorities through traffic
%   restrictions. Foreseeing the probability that \no concentrations reach
%   regulatory thresholds becomes thus a necessity both for decision makers and
%   the general public.

  % Probabilistic forecasting is a family of techniques that allow for the
  % prediction of the full expected distribution function. Thus, probabilistic
  % forecasting allows for the calculation of the future chances of exceeding the
  % protective thresholds as well as to derive cost-effectiveness analysis or to
  % detect pollution peaks.

  % In this study, we compared 10 recent probabilistic predictive models, using
  % them to predict the distribution of \no concentrations up to 60 hours in
  % advance. Quantile gradient boosted trees shows the best performance for both
  % the expected value and the forecast full distribution. We also show how this
  % approach can be used to detect pollution peaks.
  % \\\\
  % \textbf{[Required 2 line summary:]} Six methods for predicting the full
  % distribution of future \no concentrations are implemented and compared: their
  % utility is proven by forecasting 60 hour-ahead extreme pollution episodes for
  % the city of Madrid with few false positives.
\end{abstract}


\end{frontmatter}

%\linenumbers

\section{Introduction }
\label{sec:intro}

Pollution has become a worrying issue in cities due to its adverse
effects on health and the increase in pollutant concentrations, mainly
due to human activity (traffic, heating systems\ldots). In order to
take preventive steps to maintain air quality, forecasting the
evolution of pollution levels becomes a useful tool for decision
makers: detecting pollution peaks beforehand could give cities enough
time to take and communicate effective measures.

Multiple research papers have focused on this issue and have dealt
with the prediction of air quality. Bai et al. \cite{bai_air_2018}
describes the state of the art in this exercise and collects a range
of diverse solutions applied to this problem.

However, the prediction of the expected value of pollution
concentrations through point-forecasting does not provide enough
information about the likelihood of the pollutant levels reaching a
certain threshold. Indeed, we have an estimate but we usually do not
have a description of the confidence of the model nor the uncertainty
in the predictions. Therefore, it is difficult to estimate the
probability of the pollutant reaching above a certain threshold.

The reason this probability estimation is so important is because the
measures taken by cities to limit pollution (for example, limiting
traffic) impact the daily routines of citizens and prove themselves to
be quite unpopular. Therefore, local governments need to have an
estimation of the confidence in the prediction to safely engage in
those preventive measures.

% Also the use of a classifier (above or below threshold) can not 
% be used. First, we would need a classifier per threshold, but 
% more importantly, pollution peak is an extremely rare event and 
% classifiers would have to be trained with high imbalanced datasets.

As noted by Hothorn \emph{et al.} \cite{hothorn_conditional_2014}, the real
objective in a regression analysis is to find the full conditional distribution
of the target variable: in our case, the distribution of the concentration of
the pollutants. Indeed, this full distribution gives an idea of the uncertainty
of our predictions and can be useful to forecast the probability of the signal
being above a certain threshold. For example, in the city of Madrid, hourly \no
concentrations in the air are considered to be harmful from 180 $\mu g m^{-3}$.

Previous research on the same dataset has already shown the usefulness of
probabilistic forecasting for \no levels \cite{proba_aznarte}, establishing the
advantages of the approach and focusing on 1 hour-ahead predictions with a
single model (quantile random forests). We hereby extend that work by
implementing other six models (
  quantile linear regression, 
  quantile $k$-nearest neighbours, 
 quantile gradient boosted trees, 
 neural networks, 
 distributional random forests, 
 natural gradient boosting) and by using them for a wide set of
forecasting horizons (up to 60 hours). Inspiration is drawn on some of the best
approaches from the GEFCom forecasting competition
\cite{mangalova_k-nearest_2016,hong_probabilistic_2016}.

Furthermore, improving over these approaches, we also present a novel method to
apply statistical inference to the output of the models. This method aknowledges
the fact that the results show linear dependences between the predictors and the
target, which slightly benefits linear models over nonlinear ones. By combining
a linear model with nonlinear probabilistic modelling of its residuals we obtain
optimized versions of the standard models.
% This way, we are
% converting our forecasting method into a semi-parametric model.

Finally, in order to showcase one of the applications of probabilistic
forecasts, we tackle the prediction of \no peaks for the different horizons, and
compare the performance of the proposed models on this task.

\section{Probabilistic forecasting with quantile regression}
\label{sec:probForec}

The prediction from most regression models is a point estimate of the
conditional mean of a dependent variable, or response, given a set of
independent variables or predictors. However, the conditional mean does not
provide a complete summary of the distribution, so in order to estimate the
associated uncertainty, quantiles are in order. The 0.5 quantile (i.e., the
median) can serve as a measure of the center, and the 0.9 quantile marks the
value of the response below which reside the 90\% of the predicted points.
Recent advances in computing have inducted the development of regression models
for predicting given quantiles of the conditional distribution. The technique is
called quantile regression (QR) and was first proposed by Koenker in 1978
\cite{koenker_quantile_2001} based on the intuitions of the astronomer and
polymath Rudjer Boscovich in the 18th century. Elaborating from the same concept
of estimating conditional quantiles from different perspectives, several
statistical and CI models that implement this technique have been developed:
from the original linear proposal to multiple or additive regression, neural
networks, support vector machines, random forests etc.

Quantile regression has gained an increasing attention from very
different scientific disciplines \cite{yu_quantile_2003}, including
financial and economic applications \cite{ben_rejeb_financial_2016},
medical applications \cite{jang_quantile_2018}, wind power forecasting
\cite{wan_direct_2017}, electric load forecasting
\cite{lebotsa_short_2018}, environmental modelling
\cite{cade_gentle_2003} and meteorological modelling
\cite{baur_modelling_2004} (these references are just examples and the
list is not exhaustive). To our knowledge, despite its success in
other areas, quantile regression has not been applied in the framework
of air quality , with the exception of
\cite{martinezsilva_forecasting_2016}.

Thus, as we can estimate an arbitrary quantile and forecast its
values, we can also estimate the full conditional distribution, which
will entail us to the results presented in Section \ref{sec:results}.

Among the growing array of methods that allow to estimate and forecast
data-driven conditional quantiles, in this study we have chosen to
compare linear regression, $k$-nearest neighbors, random forests and
gradient boosted trees. We took the point-estimate version of those
models and converted them to their quantile or probabilistic
counterparts. We can therefore compare each model not only on the
accuracy of their point estimation but on the confidence of each
model.

We will compare the different algorithms through the RMSE, MAE and
bias for the quantile 50 and the CRPS metric for the forecast
distribution.  As described by Gneiting et
al. \cite{gneiting_probabilistic_2014}, CRPS is a measure of the
squared difference between the forecast cumulative distribution
function (CDF) and the empirical CDF of the observation.

\section{Data description and experimental design}

\subsection{Nitrogen dioxide}
\label{sec:no2}

\begin{figure}
  \centering
  \includegraphics[width=0.6\textwidth]{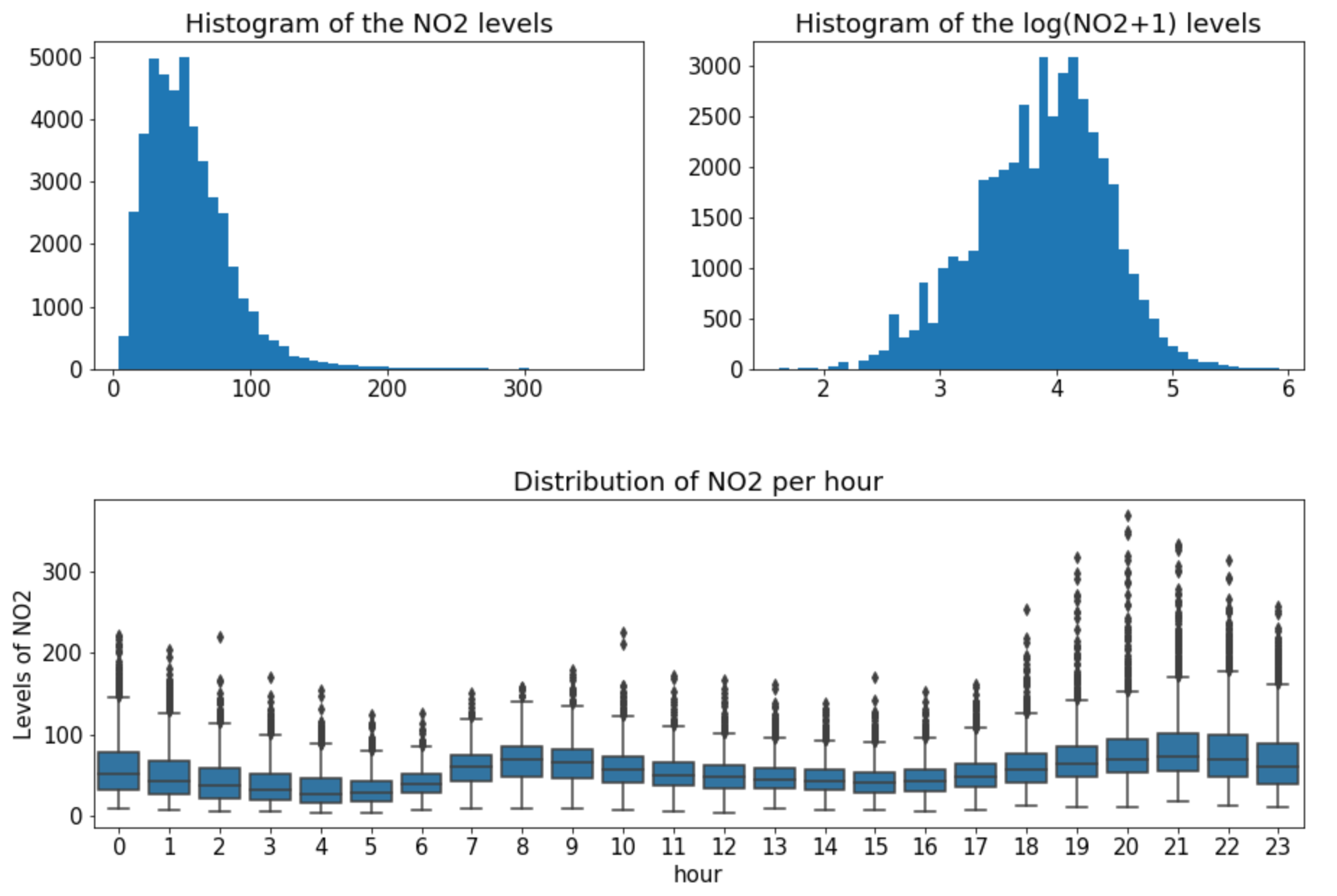}
  \caption{\label{figure:histo_variance}Distribution of logarithmic
    NO\textsubscript{2} and distribution of NO\textsubscript{2}
    per hour.}
\end{figure}

The city of Madrid has an air quality monitoring system composed by 24
stations which capture hourly data for \no.  For this study, we have
selected one of the stations with higher average leves: Escuelas
Aguirre station (code 28079008).

As we can see in \ref{figure:histo_variance}, the shape of the histogram
approaches the one from a lognormal distribution and therefore we
transformed to the logarithm of the values. This has 2 positive
effects: it reduces the tail of the distribution which will enable
better quantile estimation and it reduces the skewness of the
distribution which helps with linear models like the linear quantile
regression.

The time series for this station consists of hourly measured values of
the concentrations of \no from 01/01/2013 to
30/11/2017. These values exhibit a clear intraday pattern, in which
the higher values are located in two peaks around the morning and
evening (with highest average value around 19h) while the nightly
hours (from 00h to 05h) have lower average concentrations.  Not only
are the values higher at those hours, but also the variance is, as we
can see in figure \ref{figure:histo_variance}.
 
In order to analyze the seasonality of the signal, we extract the 5
main factors from the Fourier transform. Those correspond to the main
repetitive patterns found on the series, and can be seen clearly from
the first 3000 components. The series shows certain seasonality for
12-hour, 24-hour, one week (168-hour) and one year.  Therefore, we will
create, and use as inputs for the models, the output of periodic
functions (cosine and sine) whose frequency is equal to the ones
stated above. This will enable the machine learning models to learn
the seasonality of our time series.

As is common when forecasting with machine learning models, we exploit
the inertia of the modelled series by adding lagged variables to the
inputs. Of course, in doing so, we are limited by the horizon of the
prediction and by the 'curse of dimensionality', which implies keeping
a limited number of features as input. In our case, the inertia of the
series will be modeled by lagged values from the inmediate past (hours
before) and, based on the seasonal analysis: 1-5 hours before and
every 11-13 hours up to 9 days before.

\subsection{Ozone}

% \begin{figure}
%   \centering
%   \includegraphics[width=0.4\textwidth]{no2vso3}
%   \label{figure:no2vso3}
%   \caption{Dispersion of levels of $O_3$ vs. levels of $NO_2$.}
% \end{figure}

The same station that records Nitrogen dioxide also records the levels
of ozone (O\textsubscript{3}). It is known that ozone and Nitrogen
dioxide are related by chemical reactions occurring in the atmosphere
in the presence of sunlight, especially of the UV spectrum.  % As we can
% see in figure \ref{figure:no2vso3}, there seems to be a correlation
% between NO\textsubscript{2} levels and O\textsubscript{3}.
Thus, we will also add lagged values of O\textsubscript{3} as inputs
to our models.

\subsection{ECMWF numerical pollution prediction}
\label{sec:ecmwf-numer-poll}

The European Centre for Medium-Range Weather Forecasts (ECMWF)
implements the Copernicus Atmosphere Monitoring Service.  This service
delivers a daily production of near-real-time European air quality
analyses and forecasts with a multi-model ensemble system. Although
these forecasts are a very good starting point, 
the resolution of the model is 10km
and hence it is not expected to be capable of modelling the local
urban effects of the NO\textsubscript{2} series under study.

\subsection{Calendar Variables}
\label{sec:cal_data} 

As NO\textsubscript{2} levels are clearly be linked to human activity,
we will also flag the hours belonging to a specific type of day. Days
could be classified as bank holidays, heavy traffic days (for example,
return from holidays), school holidays\ldots We will also use as
inputs to the models past values of this variables ( 1,2 and 7 days
before ).

\subsection{Experimental Design}
\label{sec:experimental-design}

\begin{figure}
  \centering
  \includegraphics[width=0.2\textwidth]{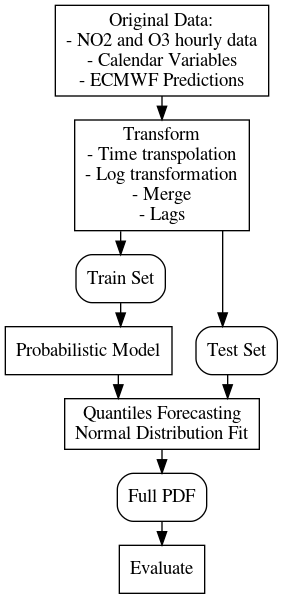}
  \caption{\label{figure:dataflow}Data flow of the experiments.}
\end{figure}

As a summary, we use the following predictors: \no
measures lagged 1-5H and every 11-13H up to 9 days before,
O\textsubscript{3} levels lagged every 24H up to 4 days before, 
calendar variables lagged 1,2 and 7 days before,
 ECMWF predictions and seasonal features extracted
from the Fourier analysis. This amounts to a total of 102 independent
variables. 

When performing the experiments, first we aligned and gathered all the
hourly time series: \no, O\textsubscript{3}, ECMWF and
calendar variables.  Then we transformed the signal levels and then we
added the lagged values and a seasonal time series with the main
periods of the NO\textsubscript{2} time series.

Once all this process is finalized, we train the following
probabilistic models: quantile random forests (QRF), $k$-nearest
neighbors (QKNN), quantile linear regression (QLR) and quantile
gradient boosting (QGB).  Figure \ref{figure:dataflow} shows the data
flow in the experimental design. All the hyperparameters of the models
have been estimated through grid search on a validation set.  
In Section \ref{sec:models}
we provide more information on each of the models.

Concerning cross-validation, there are several accepted methods to
separate the train and test set which produce correct estimations of
the error \cite{bergmeir_note_2018}: We can approximate the error of 
a model through a crossvalidated evaluation as long as
 we remove for each test set the correlated samples in the training 
 dataset. We chose a cross validation with 5 splits for time execution 
 reasons. We will always test with predictions done at
10:00, as this is the time the forecast is done in the operational
setting, as the data is first available at that time.

We want to forecast the full distribution of \no
levels for the next 60 hours and therefore we will train and evaluate
the models for each hour (60 horizons).

After forecasting the quantiles, we will fit them to a normal
distribution. Fitting a normal distribution to the predicted quantiles
and then generating the percentiles for that fitted distribution has
several advantages. It enables the calculation of more percentiles
from a small number of them.  It also helps estimating the upper tail
of the distribution, in spite of the low probability for those values.

We will evaluate the predicted 50 percentile through standard
evaluation metrics (RMSE and bias), and the predicted
distribution through the continuous ranked probability score (CRPS).

We will perform this evaluation for each of the models and each of the horizons.

\subsection{Probabilistic Models}
\label{sec:models}

As stated above, we will compare seven different probabilistic models,
which are briefly described below for reference. We will provide 
alongside the models the abreviation we will use for each of them 
throughout this article. Also we are adapting point-estimation 
algorithms to their probabilistic counterparts. This allows 
us to see the uncertainty these models have. Indeed, metrics like 
RMSE and Bias are linked to the confidence of the models for 
point estimation but the predicted CDF is much better at describing 
it. This also increases the interpretability of those models.
Therefore 
the implementations we described can have applications beyond 
this forecast exercise.

\subsubsection{Quantile linear regression (QLR)}

As shown in \cite{koenker_quantile_2001}, we can apply linear
regression with a modified cost function in order to predict the
quantiles of the dependent variable.  Given a set of vectors
$(x_i, y_i)$, in the usual point forecasting approach we are usually
interested in the prediction $\hat y(x) = \alpha_0 + \alpha_1 x$ which
minimizes the mean squared error,
\begin{equation}
  \label{eq:1}
  E = \frac{1}{n} \sum^n_i \epsilon_i =
  \frac{1}{n} \sum^n_i [ y_i - (\alpha_0 + \alpha_1 x) ]^2.
\end{equation}
This prediction is the conditional sample mean of $y$ given $x$, that
is, $\hat y(x) = \hat\alpha_0 + \hat\alpha_1 x$, or the location of
the conditional distribution. But we could be interested in estimating
the conditional median (i.e., the 0.5 quantile) instead of the mean,
in which case we should find the prediction $\hat y(x)$ which
minimizes the mean absolute error,
\begin{equation}
  \label{eq:2}
  E = \frac{1}{n} \sum^n_i \epsilon_i =
  \frac{1}{n} \sum^n_i | y_i - (\alpha_0 + \alpha_1 x) |.
\end{equation}

The fact is that, apart from the 0.5 quantile, it is possible to
estimate any other given quantile $\tau$. In that case, instead of
(\ref{eq:2}), we could minimize
\begin{equation}
  \label{eq:3}
  E= \frac{1}{n} \sum^n_i f( y_i - (\alpha_0 + \alpha_1 x))
\end{equation}
where
\begin{equation}
  \label{eq:4}
  f(y-q) = \left\{ 
    \begin{array}{l l}
      \tau (y-q) & \quad \mbox{if $y \ge q$}\\
      (1-\tau) (q-y) & \quad \mbox{if $y < q$}\\
    \end{array} \right.,
\end{equation}
with $\tau \in (0,1)$. Equation (\ref{eq:3}) represents the median
when $\tau=0.5$ and the $\tau$-th quantile in any other case.

We will train 5 linear regression models to predict 5 percentiles of
the signal. As percentiles are calculated separately, we have the risk
of quantile crossing.  We will reorder the quantiles as explained in
\cite{cross} to solve this problem.

\subsubsection{Quantile $k$-nearest neighbors (QKNN)}

We will use the probabilistic $k$-nearest neighbors algorithm as
described in \cite{quantileknnmangalova}.  This algorithm is based on
the standard $k$ nearest neighbor, where instead of calculating the
mean of the targets of the $k$ nearest points to the input, it builds
a distribution from the target of those neighbors.

\subsubsection{Quantile random forests (QRF)}

Quantile random forests create probabilistic predictions out of the
original observations. They work like the usual random forest, except
that, in each tree, leafs do not contain a single value as a
prediction but the target observations from the training set belonging
to that
leaf.

Then predictions are calculated by selecting the leafs in each tree
corresponding to the input features and combining the weighted
histograms in each tree out of the target observations in those leafs.
For more information refer to \cite{quantregforests}.

\subsubsection{Quantile Gradient Boosted Trees (QGB)}

Tree boosting \cite{friedman_greedy_2001} is a successful
machine learning technique that consist on growing trees based on the
compromise of a cost function and a regularization function. This cost
function is usually used to forecast the mean of the signal. We will
modify the cost function (im a similar way as in the quantile linear
regression) to predict the quantiles of the target. Also, we will use 
the lightgbm implementation \cite{ke_lightgbm:_2017} which provides 
lower training times and higher accuracy.

We will train 5 gradient boosted trees models to predict 5 percentiles
of the \no signal and we will solve quantile 
crossing with the technique
explained in \cite{cross}.

\subsubsection{Multilayer Perceptron}

We will also test a multilayer perceptron (MLP) 
\cite{ramchoun_multilayer_2016} with 1 hidden layer. 
Like for quantile linear regression and quantile gradient boost, 
we will modify the cost function to predict the quantiles of the target.

We will therefore as well train 5 MLP to predict 5 percentiles of the \no signal
and the whole distribution of the target.

\subsubsection{Distributional Random Forests}

Based on the idea of GAMLSS, Distributional Forests 
\cite{schlosser_distributional_2019} extend the power 
of this semiparametric method by estimating the LSS 
(location, scale and shape) through the use of random forests, 
instead of using GAM's. We will use the R implementation of this 
method from the package disttree.

\subsubsection{NGBoost}

NGBoost \cite{duan_ngboost_2019} uses the natural gradient 
instead of the standard gradient to 
build a boosted ensemble 
learner that predicts the parameters of the distribution of the target variable.
Natural Gradients are presented as offering better stability and robustness 
than normal gradients.

\subsubsection{Probabilistic forecast of linear regression residuals}

We suspect there are high linear dependencies between the
input predictors and the target. We will then experiment 
on combining 3 of the models: quantile random forest, quantile
$k$-nearest neighbor and quantile gradient boost with a linear regression.

We decided to train a linear regressor which predicts the
\no values and then use the QRF, QKNN and QGB models 
to predict the full distribution of the residuals
of that linear regression. We abbreviate both models respectively with
QRFL, QKNNL and QGBL.

\section{Results and discussion}
\label{sec:results}

\begin{figure}[tbp]
  \centering
  \includegraphics[width=0.75\textwidth]{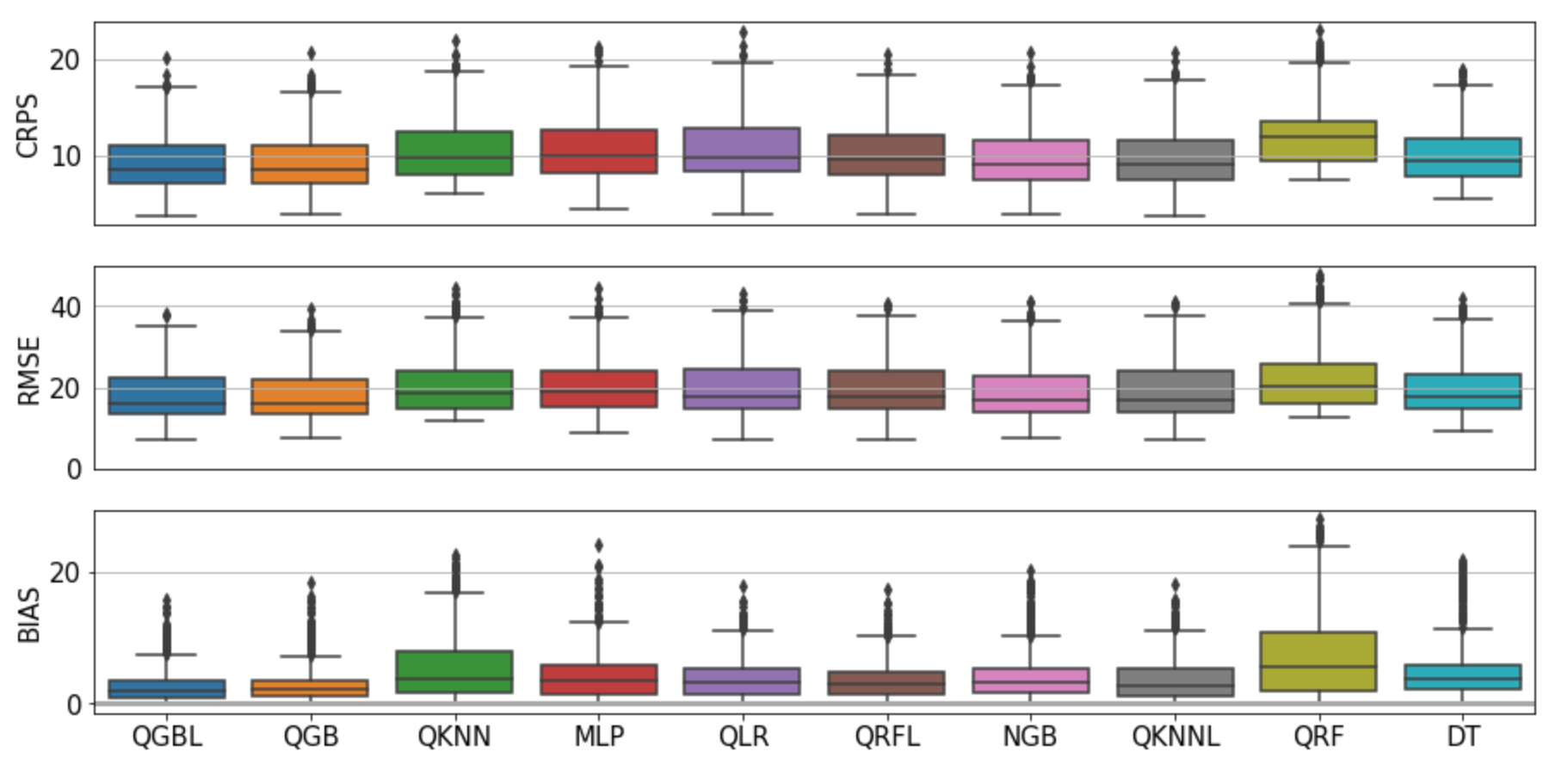}
  \caption{\label{figure:errorGraph}
    Boxplot of continuous ranked probability score, root mean squared
    error and bias of the different models for all horizons.
  }
\end{figure}

\begin{table}[tbp]
  \centering \footnotesize
  \caption{\label{tab:determ}Error measures for the proposed models.}
    \begin{tabular}{lrrlllll}
      \toprule
         method &              CRPS &              RMSE &             bias &                time &  TP &  FP &                AUC \\
        \midrule
            NGB &  $ \underset{(3.3)}{10.0} $ &  $ \underset{(7.2)}{19.4} $ &  $ \underset{(4.3)} {4.6}$ & $ \underset{(83.4)}  {997.6}$  &   4 &   3 & $ \underset{(0.41)} {0.28}$  \\
          QKNNL &  $ \underset{(3.4)}{10.0} $ &  $ \underset{(9.7)}{20.1} $ &  $ \underset{(3.7)} {3.9}$ & $ \underset{(0.6)}     {8.3}$  &  36 &  46 & $ \underset{(0.41)} {0.28}$  \\
             DT &  $ \underset{(3.1)}{10.3} $ &  $ \underset{(7.3)}{20.0} $ &  $ \underset{(5.1)} {5.3}$ & $ \underset{(17.6)}  {732.9}$  &   3 &   0 & $ \underset{(0.43)}  {0.3}$  \\
           QRFL &  $ \underset{(3.3)}{10.4} $ &  $ \underset{(9.5)}{20.6} $ &  $ \underset{(3.3)} {3.8}$ & $ \underset{(0.7)}    {10.5}$  &  24 &  55 & $ \underset{(0.4)}  {0.28}$  \\
           QKNN &  $ \underset{(3.4)}{10.7} $ &  $ \underset{(7.8)}{20.9} $ &  $ \underset{(5.2)} {5.4}$ & $ \underset{(0.6)}     {8.2}$  &   0 &   0 & $ \underset{(0.41)} {0.28}$  \\
            MLP &  $ \underset{(3.3)}{10.8} $ &  $ \underset{(6.9)}{20.6} $ &  $ \underset{(4.1)} {4.5}$ & $ \underset{(43.3)}  {232.9}$  & 131 & 219 & $ \underset{(0.41)} {0.28}$  \\
            QLR &  $ \underset{(3.6)}{10.9} $ &  $ \underset{(8.9)}{20.7} $ &  $ \underset{(3.4)} {4.0}$ & $ \underset{(0.8)}     {9.7}$  &  18 &  50 & $ \underset{(0.34)} {0.23}$  \\
            QRF &  $ \underset{(3.4)}{12.2} $ &  $ \underset{(8.6)}{22.8} $ &  $ \underset{(6.9)} {7.5}$ & $ \underset{(1.7)}    {30.8}$  &   0 &   0 & $ \underset{(0.43)}  {0.3}$  \\
            QGB &  $ \underset{(3.1)}{9.5}  $ &  $ \underset{(6.6)}{18.2} $ &  $ \underset{(3.4)} {3.3}$ & $ \underset{(8.5)}    {37.0}$  &  73 &  23 & $ \underset{(0.4)}  {0.27}$  \\
           QGBL &  $ \underset{(3.2)}{9.5}  $ &  $ \underset{(9.9)}{19.0} $ &  $ \underset{(3.0)} {2.9}$ & $ \underset{(7.4)}    {36.0}$  &  92 &  67 & $ \underset{(0.39)} {0.26}$  \\
       \bottomrule
      \end{tabular}
\end{table}

\begin{figure}
  \centering
  \includegraphics[width=0.9\textwidth]{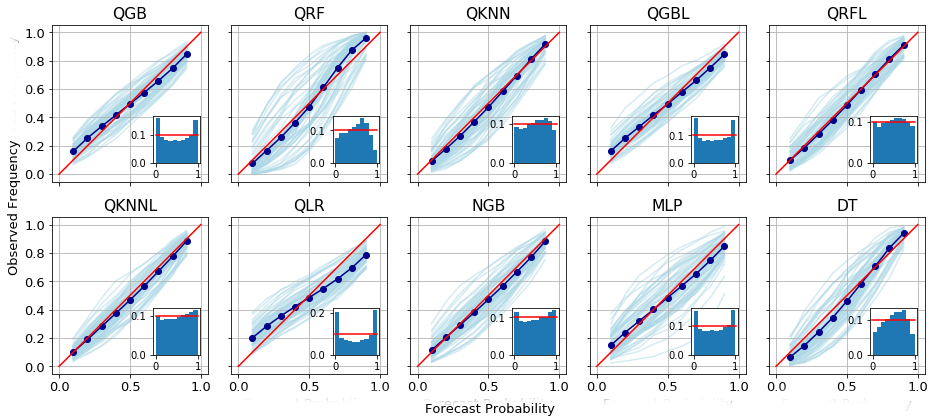}
  \caption{\label{figure:rel_sharp}Average reliability and sharpness
    of the different models across all horizons. The dim blue lines
    correspond to the different horizons. }
\end{figure}       

\begin{figure}
  \centering
  \includegraphics[width=0.49\textwidth]{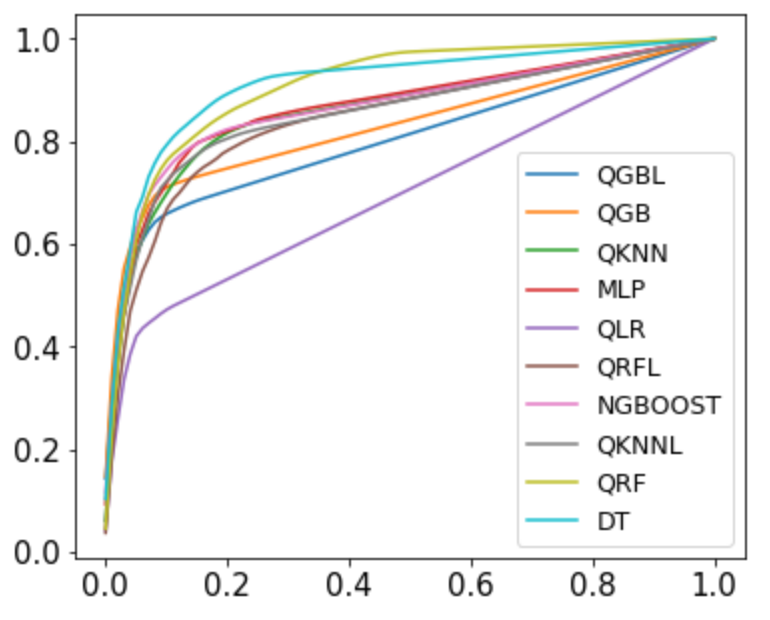}
  \caption{\label{figure:average_roc}Average ROC Curve for each 
  model. }
\end{figure}

Figure \ref{figure:errorGraph} shows the different metrics for each
model across all the horizons. First, we clearly see how 
quantile gradient
boosted trees (QGB) outperforms the other models and displays better scores
for all metrics. The additive nature of QGB is behind these
results.

Quantile random forests and quantile $k$-nearest
neighbors underperform compared to the other models, showing a bias
which is clearly higher compared to the others.  The main reason for
this is the highly linear dependence of the data. This also explains
then the good results of the linear model (QLR).  However, the linear
model underperforms when compared to other models, as it is
not able to learn the non-linear relationships between the predictors
and the target.

Regarding the models predicting the residuals 
of a linear regression, QKNNL and QRFL showcase good performance, 
which is surprising for QKNNL 
due to the simplicity 
of this model.
On the other hand, QGBL brings no real benefit compared 
to QGB, as QGB can already integrate the linear component.

Distributional Forests (DT) and Natural gradient boosted trees (NGB) 
present good results but are still outperformed by QGB. Natural gradients 
bring no benefit in this case. 

Finally, MLP has good bias results but underperforms against the other methods.

As stated earlier, the \no levels follow a lognormal
distribution and it seems that it is better modeled with a
multiplicative model (different causes multiply the level of
pollution), therefore the logarithm of the \no levels
are better forecast with an additive model. 
This is the reason why
quantile gradient boosted trees outperform 
the other models: it can naturally
add the nonlinear effects.

Table \ref{tab:determ} displays the mean and standard deviation of CRPS 
of the different models. The table shows again the good
performance of the quantile 
gradient boosted trees model for CRPS. The table also shows the training times 
of the different models. QGB appears as having a good compromise 
between low training times and metrics. QKNNL offers good performance with really low 
training times.

For probabilistic models, CRPS is a good summary of the performance of
the models. Notwithstanding, the reliability and sharpness graphs are
known to be useful at estimating how the observed values are
positioned in the distributions.  Figure \ref{figure:rel_sharp}
features the reliability and sharpness of the different models.

The sharpness curve shows that 
QRF
and QKNN tend to underestimate the levels of
\no. This is clear in the sharpness curve for the
upper percentiles. The observed values of those percentiles are
usually noticeably higher than the forecast values. This means that,
too many times, the observed values are in the upper percentiles of
the distribution. This is also a consequence of the bias of both
models.

For QRF and DT, we see few 
observations on the sides of the predicted distributions. This means 
the variance of the predicted variance is too high and the predictions 
have too much uncertainty.

QGB and QLR forecast
distributions seem more balanced but display high values at both sides
of the sharpness curve.  If we consider the distribution to be a
gaussian distribution, this means the forecast distributions tend to
have a too small standard deviation and are too narrow. Therefore, the
forecast probability of some levels of \no is too
small compared to the observed one. QGBL and MLP also display this behaviour.
As we are interested in predicting peaks of pollution, 
this is not detrimental.

%\subsection{Advanced Non Parametric Statistical tests on the results}

In order to compare the result of the 10 methods, we will use advanced 
non parametric tests as described by García et al. \cite{garcia_advanced_2010} . 
The Quade test will calculate the rank of all the algorithms and 
then determine if they are significantly differently from the mean rank. 
We have 60 different datasets, each one corresponding to a horizon 
where we have performed 
a 5-fold crossvalidation and kept the mean of the results.
The rank is calculated not only based on the results in the different 
datasets but also on the importance of the dataset 
(based on the variance of results). 
The result of the Quade Test rejects the null hypothesis that the methods have 
similar performance and we can use the ranks to classify the different methods. 
Table \ref{tab:quade} shows the results of the Quade test.

\begin{table}[tbp]
  \centering \footnotesize
  \caption{\label{tab:quade}Average Rankings of the algorithms (Quade)}
  \begin{tabular}{c|c}
    Algorithm&Ranking\\
    \hline
    DT&4.864155251141552\\
    MLP&4.289954337899544\\
    NGB&6.81544901065449\\
    QGB&9.407153609071533\\
    QGBL&8.90677321156773\\
    QKNN&3.10806697108067\\
    QKNNL&6.51255707762557\\
    QLR&4.276255707762556\\
    QRF&1.4448249619482494\\
    QRFL&5.3748097412480975\\
    \end{tabular}
\end{table}

\subsection{Forecasting \no concentration peaks}

One of the applications of probabilistic forecasts is the estimation
of the probability of the signal being above a certain threshold. In
the case of \no forecasting, this is especially useful when normative
limits are established by authorities.

We evaluated the proposed models to check how good are they to detect
those peaks.  In Madrid, municipal regulations fix a threshold of 180
$\mu gm^{-3}$ to activate the first level of restrictive measures on
traffic. Thus, we consider that a \no peak is forecast when the
probability of reaching above this limit is higher than 50\%.  It is
important to note that this normative threshold lies around the 99.7
quantile of the \no distribution, and thus it is safe to consider
peaks as rare events.
This fact of course implies that the possibility of using a classifier
to predict peaks, with classes above or below the threshold, is not
straightforward. Firstly, we would need a classifier per threshold,
but more importantly, it would suffer from the fact that the training
dataset would be highly imbalanced. A probabilistic model can learn from the
predicted distribution even if the peak level has not been 
reached. 

As noted before, there is usually a high cost on performing preventive
measures against pollution. This has some consequences for the
evaluation of our models. We must minimize the number of false
positives and false negatives and thus secondary metrics like the ROC
curve are not directly applicable as in this case, depending on the
take of the decision makers, the cost of a false positive might be
much higher than the cost of a false negative. 
Table \ref{tab:determ} shows the true positives and false positives
of each model. We see again the superior results of QGB against the other models as it 
has a good ratio of true positives against false positives.
Finally, Area under curve 
"AUC" seems to not be a good metric to compare the different models as the differences are 
not that high.

\section{Conclusions}
\label{sec:concl}

After extracting and processing the data from one of the pollution stations in
Madrid, we have compared 10 different models to build a probabilistic forecast
of the levels of \no for up to 60 hours into the future. We have evaluated our
models through the forecast quantile 50 and the full forecast distribution.

We have observed a linear dependence between the target and the features. For
this reason, the linear quantile regression has performed well compared to
random forests and $k$-nearest neighbors. However, the multiplicative nature of
the levels of \no and the nonlinear dependence between target and features have
lead to better results for the gradient boosted trees which has outperformed all
the other models in all metrics.

However, we have shown how quantile random forest and quantile $k$-nearest
neighbors could be used to improve the results of a linear model when nested to
model the full distribution of the residuals of a linear regression. Those
models, specially the $k$-nearest neighbor are easier to train, so they become
worthy alternatives to the gradient boosted trees.

In sum, we have tested 10 alternative models to produce probabilistic forecasts
for \no, and we have compared them through different metrics. Also, we have
established how quantile gradient boosted trees are able to detect pollution
peaks beforehand with few false positives.

% In the future, our work could be enhanced by improving the quality of
% the probabilistic forecast either by using new models (neural
% networks) or extracting better features as input of our models.

%\section*{References}
\bibliographystyle{abbrv}

\end{document}